\documentclass[prd,aps,preprint,epsfig,floats,showaps]{revtex4}
\usepackage{amsmath}
\usepackage{amsfonts}
\usepackage{graphicx}
\usepackage{epsfig}

\begin{document}
\newcommand{\vect}[1]{\overrightarrow{#1}}
\newcommand{\smbox}[1]{\mbox{\scriptsize #1}}
\newcommand{\tbox}[1]{\mbox{\tiny #1}}
\newcommand{\slh}[1]{\displaystyle{\not}#1}
\newcommand{\gL}{g_{\mbox{\tiny{L}}}}
\newcommand{\gR}{g_{\mbox{\tiny{R}}}}
\newcommand{\gY}{g_{\mbox{\tiny{Y}}}}
\newcommand{\YL}{Y_{\mbox{\tiny{L}}}}
\newcommand{\YR}{Y_{\mbox{\tiny{R}}}}
\newcommand{\PL}{P_{\tbox{L}}}
\newcommand{\PR}{P_{\tbox{R}}}
\newcommand{\Neu}{\mathcal{N}}
\newcommand{\rt}{\sqrt{2}}
\newcommand{\hc}{\ensuremath{\mathrm{h.c.}}}
\preprint{UMD-PP-07-004}
\preprint{MSU-HEP-07-08-28}
\title{\Large\bf Pseudo-Dirac bino dark matter}
\author{\bf Ken Hsieh}
\email[email: ]{kenhsieh@pa.msu.edu}

\affiliation{Department of Physics, University of Maryland, College Park, MD 20742, USA\\
Department of Physics, Michigan State University, East Lansing, MI 48824, USA}

\date{\today}

\begin{abstract}
While the bino-dominated lightest neutralino of the minimal
supersymmetric Standard Model (MSSM) is
an interesting and widely-studied
candidate of the dark matter, the $p$-wave suppression
of its annihilation cross section requires fine-tunings of the MSSM spectra
to be consistent with Wilkinson Microwave Anisotropy Probe (WMAP) observations.
We propose a pseudo-Dirac bino
that arises in theories with D-type supersymmetry-breaking
as an intriguing alternative candidate of dark matter.
The pseudo-Dirac nature of the bino gives a natural mechanism
of enhanced co-annihilation because these two states are degenerate
in the absence of electroweak symmetry breaking.
In addition, the lightest state can be consistent with
limits of direct detection experiments because of
the lack of vector interactions, as with the case
of the MSSM bino.
\end{abstract}

\maketitle

\section{INTRODUCTION}
The existence of dark matter is one of the direct evidences of
physics beyond the standard model (SM), and supersymmetry (SUSY) is a strong
candidate of such new physics.
One the many virtues of the minimal supersymmetric Standard Model (MSSM)
is that the lightest superpartner can serve as dark matter of the Universe.
The lightest neutralino of the MSSM is an interesting
and widely-explored candidate
of dark matter
(\cite{Griest:1988ma}\cite{Griest:1989zh}\cite{Jungman:1995df}\cite{Bottino:1996eu}\cite{Choi:2000kh}\cite{Nihei:2002ij}\cite{Nihei:2004bc}
and references therein).
Because of electroweak symmetry breaking (EWSB), the lightest neutralino, $\chi_1^0$, is linear combination
of the bino ($\lambda_1$), the wino ($\lambda_2$), and the Higgsinos ($\tilde{H}_{u,d}$).
Expressing $\chi_1^0$ as
\begin{align}
\chi_1^0=Z_{11}\lambda_1+Z_{12}\lambda_2+Z_{13}\tilde{H}_d+Z_{14}\tilde{H}_u,
\end{align}
where $Z_{ij}$ are elements of the transformation matrix that diagonalizes the
MSSM neutralino mass matrix, there are several important limits of
interest. In the Higgsino-dominated ($Z_{13}\simeq Z_{14}\gg
Z_{11}, Z_{12}$) and the wino-dominated ($Z_{12}\gg
Z_{11},Z_{13},Z_{14}$) limits, $\chi_1^0$ is typically nearly
degenerate with either the charged Higgsinos and/or the charged
winos, and the efficient self- and co-annihilation processes leads
to a relic density that is less than the WMAP observation \cite{wmap}
\begin{align}
\Omega h^2= 0.1195\pm 0.0094.
\label{eq:wmap-bound}
\end{align}
On
the other hand, in the bino-dominated limit ($Z_{11}\gg
Z_{12},Z_{13},Z_{14}$), the annihilation cross section is $p$-wave
suppressed, and bino relic density is typically higher than the
WMAP observation.

For the relic density of $\chi_1^0$ to be consistent with current
observations, there typically requires fine-tunings of the MSSM spectra
such that one or more of the following occurs \cite{EllisKingRoberts}:
\begin{itemize}
\item Enhanced co-annihilation between $\chi_1^0$ and another superpartner,
typically the s-tau ($\tilde{\tau}$), when these two states are tuned to be
nearly degenerate on the order of 1\%.
\item Enhanced $s$-channel resonance in the $\chi_1^0$ annihilation cross section
when the mass of one of the Higgs bosons is tuned to be close to twice the mass of $\chi_1^0$.
\item Enhanced annihilation cross section from the wino/Higgsino mixture of $\chi_1^0$
and enhanced co-annihilation of the charginos when either the mass of the winos (both neutral and charged)
or the Higgsinos (both neutral and charged) is tuned to be close to the bino mass
\cite{Arkani-Hamed:2006mb}.
\end{itemize}

The root of the problem is the $p$-wave suppression of the bino annihilation cross section
due to the Majorana nature of the bino.  On the other hand, Dirac particles carrying
$SU(2)_L$ or $U(1)_Y$ quantum numbers
such as the Kaluza-Klein (KK) neutrino
of the minimal universal extra dimension model
\cite{acd}\cite{Servant:2002hb}
are typically ruled out as dark matter by
direct detection experiments such the Cryogenic Dark Matter Search (CDMS II)\cite{cdms}
and the XENON10 Dark Matter Experiment \cite{Angle:2007uj}.
(For models where Dirac particles serve as viable dark matter,
see Refs.~\cite{KH}\cite{Belanger:2007dx}.)
We would like a natural mechanism that enhances the annihilation
cross section, and at the same time, is consistent with the bounds
of direct detection. The pseudo-Dirac bino is one such example.
Pseudo-Dirac bino may arise in models of D-type SUSY-breaking.
However, existing models \cite{DGMSB2} predict a heavy pseudo-Dirac bino with
masses at least of the order of 1 TeV.
In this paper, we consider pseudo-Dirac bino as a candidate of dark matter.
Without effects of EWSB, the bino is a Dirac particle whose
annihilation cross section is not $p$-wave suppressed and can naturally
lead to observed relic density.  When
EWSB effects are considered, the Dirac bino splits into
two nearly-degenerate Majorana states, and the annihilation
of the lightest state is enhanced by co-annihilation between
the these two nearly-degenerate bino states.
On the other hand, the masses of these two states are separated
by a few GeV's while the scale of momentum transfer in
direct detection experiments is of the order of keV's.
Therefore, the direct detection experiments is only sensitive to
the lightest,
Majorana, state whose cross section with nuclei is suppressed
due to the lack of vector-current interactions.
It is worth pointing
out that this mechanism of suppressing the rates of direct
detection operates as long as the splitting is larger than 10s of keV's,
and is not limited to the splitting of a few GeV's (which happens
to be our case here).
For a similar idea involving the sneutrino as dark matter,
see Reference \cite{Hall:1997ah}.

In this paper we take a phenomenological approach, without appealing to a complete
framework, and perform a simplified analysis
of the relic density and direct detection rates of
pseudo-Dirac bino dark matter.  In Section \ref{sec:DGMSB},
we describe the relevant ingredients of D-type SUSY-breaking
that lead to the pseudo-Dirac
bino as dark matter.  In Section \ref{sec:DM}, we calculate
the relic density and direct detection rates of
pure-Dirac and pseudo-Dirac bino dark matter, and compare
the results to those in MSSM.
Finally, we summarize our results in Section \ref{sec:Conc}.

\section{D-TYPE GUAGE MEDIATED SUPERSYMMETRY BREAKING MODEL}
\label{sec:DGMSB}
We assume that SUSY-breaking originates in a hidden sector that contains a gauged $U(1)_X$ group that develops
a non-zero $\langle D_X\rangle$, as well as non-zero $\langle F_Y\rangle$ for some field(s) $Y$
that may or may not be charged under the $U(1)_X$ group (but neutral under SM gauge group).
In general, both $\langle D_X\rangle$ and $\langle F_Y\rangle$ are communicated to the visible MSSM sector.
Upon integrating out the messengers at the mass scale $M$, the Majorana gaugino masses are generated
through the effective operator
\begin{align}
\mathcal{L}\sim\int d^2\theta \frac{Y}{M}W_{\alpha}W^{\alpha}+\mbox{h.c.},
\label{eq:MajoranaMass}
\end{align}
where $W_{\alpha}$ is the chiral superfield containing the MSSM gaugino and gauge bosons,
while MSSM scalar masses are generated through the effective operator
\begin{align}
\mathcal{L}\sim\int d^4\theta \frac{Y^{\dag}Y}{M^2}Q^{\dag}Q.
\label{eq:ScalarMass}
\end{align}

In our phenomenological approach, we assume that $Y$ is charged under $U(1)_X$, and thus the effective
operator of Eq.~(\ref{eq:MajoranaMass}) is not generated, while MSSM scalar soft masses
are still generated through the operator in Eq.~(\ref{eq:ScalarMass}).
This can be achieved, for example, by the charge assignments of the messengers
and the hidden sector particle content under the $U(1)_X$ gauge group.
Although this does not solve the flavor problem of the MSSM, we will take this as our starting point
for the purpose of discussing pseudo-Dirac bino as dark matter.
%
With the above assumptions of SUSY-breaking, the gauginos of the MSSM receive Dirac masses rather than Majorana masses.
As a Dirac fermion contains
more degrees of freedom than a Majorana fermion, additional fermionic states (the gaugino partners,
denoted by $\xi$)
that transform as adjoints of the SM group
must be introduced.  Supersymmetry (SUSY) then requires additional bosonic states (the s-gaugino, denoted
by $\eta$)
that also transform as adjoints of the SM group.
The effective operator obtained by integrating out the messengers that gives a Dirac gaugino mass
is
\begin{align}
\mathcal{L}\sim \int d^2\theta \frac{X_{\alpha}}{M} \mbox{Tr}\left[W^{\alpha}\Xi\right]+\mbox{h.c.},
\label{eq:DiracGauginoMass}
\end{align}
where $M$ is the mass scale of the messengers, and $\Xi$ is a chiral superfield containing $\eta$ and $\xi$.
We can forbid Majorana masses for the gaugino partners of the form
\begin{align}
\mathcal{L}\sim \int d^2\theta M\mbox{Tr}\left[\Xi\Xi\right]+\mbox{h.c.},
\label{eq:MajorGauginoMass}
\end{align}
by $U(1)_R$ symmetry that assigns the vector superfields $W_{\alpha}$ and $X_{\alpha}$ to
have $U(1)_R$ charge of +1 and $\Xi$ to have a zero $U(1)_R$ charge.
Since the superpotential needs to have a $U(1)_R$ charge of +2, the effective operator
of Eq.~(\ref{eq:DiracGauginoMass}) is allowed by $U(1)_R$, while the operator of
Eq.~(\ref{eq:MajorGauginoMass}) is forbidden.
We assume that the
Dirac gaugino masses and the soft scalar masses (for both the MSSM superpartners and the s-gaugino)
are all of the same scale of the order 1 TeV.
Since the gaugino partners are odd under matter-parity, an immediate interesting consequence of
D-type SUSY-breaking models is that
the s-gauginos are
even under matter-parity and could be singly produced at the Large Hadron Collider (LHC).

Dirac gaugino masses are super-soft, and do not enter the renormalization group equations (RGEs)
of the scalar soft masses.  Ignoring all Yukawa couplings except for the top Yukawa coupling,
the RGEs of all the soft s-fermion masses (except for the s-top masses) vanish at one-loop.
The dominant two-loop contributions to the RGEs involve $m^2_{\eta}$ and are negative.
Thus, if the soft masses are unified at the grand unified theory (GUT) scale, we would have
a compact (compared to the typical models of SUSY-breaking such as gauge- and anomaly-mediated SUSY breaking)
and inverted spectra with sleptons heavier than the squarks.
In particular, the s-top would be the lightest
sfermion and its mass can be approach current experimental bounds ($m_{\tilde{t}}>$ 300 GeV)
without s-leptons violating current experimental bounds ($m_{\tilde{l}}>$ 100 GeV).
Such spectra of D-type SUSY-breaking are very distinct from the typical MSSM
spectra obtained by gauge-mediated supersymmetry breaking and other generic models of SUSY-breaking.

There are no trilinear soft terms in models of D-type SUSY-breaking,
and the s-top masses can be as light as 400 GeV.  While this
may potentially solve the little hierarchy problem, where large radiative
corrections to the soft Higgs mass $m^2_{H_u}$ requires a fine-tuning
of a few percent to achieve successful EWSB, large $m_{\tilde{t}}$ and/or $A_t$ are needed for the mass of the
lightest CP-even boson to satisfy the CERN LEP bounds \cite{lep2} of $m_{h}>114.4$ GeV.
Since we do not offer a complete model, we here give only a few remarks
about EWSB with D-type SUSY-breaking.

%
%
%
%
One possibility of having successful EWSB is to extend the Higgs sector
with an additional singlet chiral
superfield, $S$, with the superpotential
\begin{align}
\Delta W=\lambda S H_u H_d +\frac{\kappa}{3}S^3,
\label{eq:NMSSM-W}
\end{align}
that replaces the $\Delta W=\mu H_u H_d$ term in the MSSM superpotential.
While this superpotential of Eq.~(\ref{eq:NMSSM-W}) is same
as that of next-to-minimal supersymmetric Standard Model (NMSSM),
unlike the typical NMSSM scenarios, we do not have trilinear SUSY-breaking terms
in the potential.
Instead, we include terms
\begin{align}
\Delta\mathcal{L}=B_H(H_uH_d+\hc)+B_S(S^2+\hc),
\end{align}
and still achieve successful EWSB with a lightest CP-even boson that satisfies
the LEP2 bounds.
It is worth emphasizing that, unless the fermionic component of $S$,
the singletino ($\tilde{s}$), mixes significantly with the bino, our following analysis does not
depend on the existence the chiral superfield $S$.

\section{PSEUDO-DIRAC BINO DARK MATTER}
\label{sec:DM}
The mass matrix of the neutral neutralino in a D-type SUSY-breaking scenario
in the basis
$(\lambda_1\ \ \xi_1\ \ \lambda_2\ \ \xi_2\ \ \tilde{H}_d\ \ \tilde{H}_u\ \ \tilde{s})$
is
\begin{align}
\mathcal{M}_0=
\begin{pmatrix}
0 & M_1 & 0 & 0 & -\frac{\gY}{2}v_d & \frac{\gY}{2}v_u & 0 \\
M_1 & 0 & 0 & 0 & 0 & 0 & 0 \\
0 & 0 & 0 & M_2 & \frac{g_2}{2}v_d & -\frac{g_2}{2}v_u & 0 \\
0 & 0 & M_2 & 0 & 0 & 0 & 0 \\
-\frac{\gY}{2}v_d & 0 & \frac{g_2}{2}v_d & 0 & 0 & \mu & \frac{\lambda}{2} v_u \\
\frac{\gY}{2}v_u & 0 & -\frac{g_2}{2}v_u & 0 & \mu & 0 & \frac{\lambda}{2} v_d \\
0 & 0 & 0 & 0 & \frac{\lambda}{2}v_u & \frac{\lambda}{2}v_d & 2\frac{\kappa}{\lambda}\mu
\end{pmatrix},
\label{eq:MassMatrix}
\end{align}
where $M_{1,2}$ are the Dirac bino and wino mass, respectively.
The gauge couplings of $U(1)_Y$ and $SU(2)_L$ SM gauge groups are denoted by
$\gY$ and $g_2$, respectively,
$v_{u,d}=\sqrt{2}^{-1}\langle H_{u,d}\rangle$, and $\mu=\sqrt{2}^{-1}\lambda\langle S\rangle$.
Since we are interested in the bino-dominated limit, we will assume that
$m_2, \mu > m_1$.  To simplify our analysis, we will also make these following three assumptions.
\begin{itemize}
\item First, the mass of the lightest bino state is smaller than
the mass of the $W$-boson, $M_W$, so the only possible
annihilation products are fermion-antifermion pairs.
While the annihilation channels
into the gauge and the Higgs bosons can be important for wino- and Higgsino-dominated
$\chi_1^0$ of the MSSM, the fermion-antifermion annihilation channels dominate
the total annihilation cross section in
the bino-dominated $\chi_1^0$ even when the gauge boson annihilation channels are kinematically
allowed \cite{Griest:1989zh}.  For the D-type SUSY-breaking scenario, we will simply assume this and postpone the verification
of this assumption in a later study.
\item Second, we assume that $M_1$, $M_2$ and $\mu$ are all positive.  While the
relative signs and phases of these parameters are important when making a detailed
study, we will assume this simple case.
\item Third, the matrix $\mathcal{M}_0$ has the hierarchy
\begin{align}
\mu \gg m_1\sim m_2\sim v_{u,d},
\end{align}
so we can expand in $\mu^{-1}$ and keep the lowest terms.  However,
we do not assume that $m_1$ and $m_2$ are nearly-degenerate,
so there are no co-annihilation contributions from the charged winos.
\end{itemize}

With these three assumptions, we first compute the relic density
in the limit of pure Dirac bino ($\mu\rightarrow\infty$), and then compute the corrections
induced by EWSB to first-order in the effects of EWSB and $\mu^{-1}$.
We then compute the direct detection cross section of pure- and pseudo-Dirac bino
to the same order.

\subsection{Relic density in the pure Dirac bino limit}
In the limit of large $\mu$, the Higgsinos and the singletino decouple and
the lightest neutralino state is a pure Dirac bino.
In terms of two-component Weyl spinors, we have the following Lagrangian
of the Dirac bino mass and bino-fermion-sfermion interactions
\begin{align}
\Delta\mathcal{L}=
-\sqrt{2}\gY \YL
(\lambda_1 q_{\tbox{L}} \tilde{q}_{\tbox{L}}^{\ast}
+\lambda_1^{\dag} q^{\dag}_{\tbox{L}} \tilde{q}_{\tbox{L}})
-
\sqrt{2}\gY \YR
(\lambda_1 \overline{q_{\tbox{R}}} \tilde{q}_{\tbox{R}}^{\ast}
+\lambda_1^{\dag} \overline{ q_{\tbox{R}} }^{\dag} \tilde{q}_{\tbox{R}})
-M_1(\lambda_1\chi_1+\lambda_1^{\dag}\chi_1^{\dag}),
\label{eq:original-gauge-int}
\end{align}
where $q_{\tbox{L}}$ and $\overline{q_{\tbox{R}}}$ are two-component SM fermion with hypercharge
$\YL$ and $\YR$, respectively.
We define the Dirac spinors
\begin{align}
Q=\begin{pmatrix} q_{\tbox{L}} \\ \overline{q_{\tbox{R}}}^{\dag} \end{pmatrix},\quad
D=\begin{pmatrix} \lambda_1 \\ \chi_1^{\dag} \end{pmatrix},\quad
D^{\prime}=C\overline{D}^T=\begin{pmatrix} \chi_1 \\ \lambda_1^{\dag} \end{pmatrix},
\end{align}
and the projection operators
\begin{align}
\PL=\frac{1}{2}(1-\gamma_5)=\begin{pmatrix}1 & 0 \\ 0 & 0 \end{pmatrix},\quad
\PR=\frac{1}{2}(1+\gamma_5)=\begin{pmatrix}0 & 0 \\ 0 & 1 \end{pmatrix}.
\end{align}
We can then rewrite the Lagrangian in terms of the Dirac spinors
\begin{align}
\mathcal{L}=
-\sqrt{2}\gY\YL(\overline{D}^{\prime}\PL Q \tilde{q}_{\tbox{L}}^{\ast}
+\overline{Q}\PR D^{\prime} \tilde{q}_L)
-\sqrt{2}\gY\YR(\overline{D} \PR Q \tilde{q}_{\tbox{R}}^{\ast}
+\overline{Q} \PL D \tilde{q}_{\tbox{R}})
-M_1\overline{D}D.
\label{eq:D-gauge-int}
\end{align}
Integrating out the sfermions we obtain
the effective four-fermion interactions
\begin{align}
\mathcal{L}_{\tbox{eff}}
=
\frac{2\gY^2\YL^2}{M^2_{\tilde{q}_{\tbox{L}}}}
(\overline{D}^{\prime}\PL Q)(\overline{Q}P_R D^{\prime})
+
\frac{2\gY^2\YR^2}{M^2_{\tilde{q}_{\tbox{R}}}}
(\overline{D} \PR Q)(\overline{Q} \PL D).
\label{eq:D-eff}
\end{align}
Applying Fierz transformation, we obtain
\begin{align}
\mathcal{L}_{\tbox{eff}}
=
\frac{\gY^2\YL^2}{M^2_{\tilde{q}_{\tbox{L}}}}
(\overline{D}^{\prime}\gamma^{\mu}\PL D^{\prime})(\overline{Q}\gamma_{\mu} \PR Q)
+
\frac{\gY^2\YR^2}{M^2_{\tilde{q}_{\tbox{R}}}}
(\overline{D}\gamma^{\mu} \PR D)(\overline{Q}\gamma_{\mu} \PL Q),
\label{eq:D-eff2}
\end{align}
which will be useful when we compute the direct-detection rate in the limit of
pure Dirac bino dark matter.

From the effective interactions of
Eqs.~(\ref{eq:D-eff}) and (\ref{eq:D-eff2}),
we have the thermal-averaged
annihilation cross section
\begin{align}
\langle\sigma(\overline{D}D\rightarrow \overline{f}f)v\rangle=
\frac{\gY^4 M_1^2}{8\pi}\sum\limits_{f}\frac{N_f Y_f^4}{M^4_{\tilde{f}}}
\left(1+\mathcal{O}\left(\frac{T}{M_1}\right)\right),
\label{eq:D-sig}
\end{align}
where $v$ is the relative velocity of the annihilating binos and
the summation sums over all the fermions of the SM except for the top quark, $N_i$
is the color factor ($N=3$ for quarks and $N=1$ for leptons), and $T$ is the temperature
of Dirac bino.
Since this annihilation cross section is not $p$-wave suppressed,
it is a good approximation to keep the leading, temperature-independent,
contribution, as we have done here.
The relic density of the pure Dirac bino is then given by \cite{KolbTurner}
\begin{align}
\Omega h^2=2\frac{x_F}{\sqrt{g_{\ast}}}
\frac{8.7\times 10^{-11}\ \mbox{GeV}^{-2} }{ \langle\sigma(\overline{D}D\rightarrow \overline{f}f) v\rangle},
\label{eq:Omega-DB}
\end{align}
where $g_{\ast}=96$ is the number of relativistic degrees of freedom at
the freeze-out temperature $T_F$, and $x_F=M_1/T_F$.  Also, in Eq.~(\ref{eq:Omega-DB}),
we have included a factor of 2 to account for the relic density of both the particle
\emph{and} the antiparticle, as explained in the Appendix of Ref.~\cite{Srednicki:1988ce}.
In general, the freeze-out temperature of species $A$ with mass
$M_A$ is
given by iteratively solving the formula
\begin{align}
x_F=\ln\left( \frac{5}{4} \sqrt{\frac{45}{8}} \frac{d_A}{2\pi^3}
\frac{ M_A M_{\tbox{Pl}} }{ \sqrt{g_{\ast}x_F} }
\langle\sigma(AA\rightarrow XX) v_F\rangle
\right),
\label{eq:xF-D}
\end{align}
where $d_A$ is the degrees of freedom of $A$, and $\langle\sigma(AA\rightarrow XX) v_F\rangle$
is the thermal-averaged cross section evaluated at the freeze-out temperature
\begin{align}
\langle\sigma(AA\rightarrow XX) v_F\rangle
\equiv
\langle\sigma(AA\rightarrow XX) v\rangle|_{T\rightarrow T_F}.
\label{eq:freeze-out-temp}
\end{align}
%
%
%
%

As a comparison, the relic density of a pure Majorana bino in the MSSM is
(see Reference \cite{Arkani-Hamed:2006mb}, for example)
\begin{align}
\Omega h^2\simeq 2\frac{x_F}{\sqrt{g_{\ast}}}
\frac{8.7\times 10^{-11}\ \mbox{GeV}^{-2} }
{ \langle\sigma(\tilde{B}\tilde{B}\rightarrow \overline{f}f) v_F\rangle},
\label{eq:Omega-MB}
\end{align}
where
\begin{align}
\langle\sigma_{\tilde{B}\tilde{B}} v_F\rangle
=\frac{\gY^4}{2\pi}\sum\limits_f
N_f
Y_f^4\frac{r_f(1+r_f^2)}{M^2_{\tilde{f}} (1+r_f)^4 x_F},\quad\mbox{with}\quad
r_f\equiv \frac{M^2_1}{M^2_{\tilde{f}}},
\label{eq:M-Bino-anni}
\end{align}
is the thermal-averaged annihilation cross section evaluated at
the freeze-out temperature $T_F$, which can
be solved from Eq.~(\ref{eq:xF-D}).

\begin{figure}[htb]
\begin{center}
\includegraphics[width=4in]{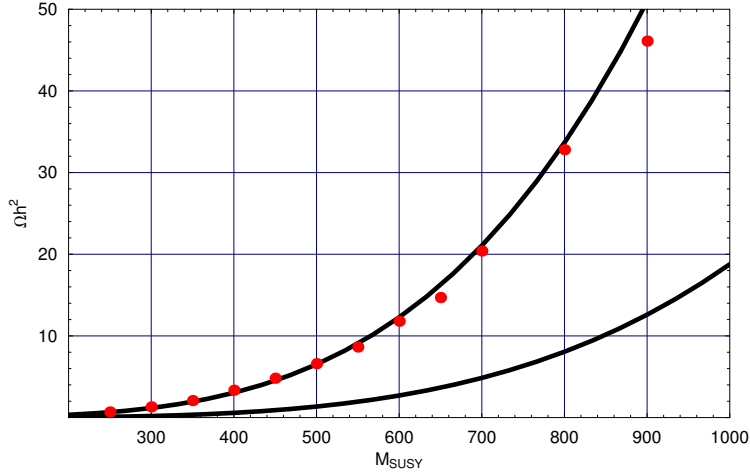}
\caption{The relic densities $\Omega h^2$ of pure Dirac (lower line) and Majorana (upper line) bino
as a function of a common sfermion mass $M_{\tbox{SUSY}}$.  The dots on top of the upper line are
computed using MicrOMEGAs 2.0 with spectra whose $\chi_1^0$ is mostly the Majorana bino.}
\label{fig:DBinoPlot}
\end{center}
\end{figure}

In Figure \ref{fig:DBinoPlot}, we plot Eqs.~(\ref{eq:Omega-DB}) and (\ref{eq:Omega-MB})
as functions of a common scalar soft mass $M_{\tbox{SUSY}}$,
as well as the relic density calculated by MicrOMEGAs 2.0 \cite{micromegas}
as checks for sample spectra that approach the bino-dominated limit.
Although neither results
are consistent with the WMAP observational bounds
of Eq.~(\ref{eq:wmap-bound}), we see that the relic
density of a pure Dirac bino
is smaller by roughly a factor of 4 compared
to that of the Majorana bino, and there may be less fine tuning
in the D-type SUSY breaking models than the MSSM to obtain the observed
relic density of dark matter.

\subsection{Relic density of pseudo-Dirac bino}
Because of EWSB contributions, the D-type SUSY-breaking spectra has a pseudo-Dirac bino consisting
of two nearly-degenerate bino states when $\mu\gg M_1, M_2$.
Expanding the effective bino mass matrix to order $\mathcal{O}(\mu^{-1})$, we have
\begin{align}
\mathcal{M}_{\tbox{bino}}=
\begin{pmatrix} \gY^2\frac{v_uv_d}{2\mu} & M_1 \\ M_1 &  0 \end{pmatrix},
\label{eq:eff-bino-matrix}
\end{align}
giving the mass eigenstates
\begin{align}
\chi_{1,2}^0&=\frac{1}{\rt}(\lambda_1\mp\xi_1),
\end{align}
with masses
\begin{align}
|M_{\chi_{1,2}^0}|=M_1\mp\frac{\gY^2 v_u v_d}{4\mu},
\end{align}
where we have used the assumption that $M_1>0$.
The gauge interactions of Eq.~(\ref{eq:D-gauge-int}) can now
be written as
\begin{align}
\Delta\mathcal{L}
&=
-\gY \YL
(\chi_1 q_{\tbox{L}} \tilde{q}_{\tbox{L}}^{\ast}
+\chi_1^{\dag} q^{\dag}_{\tbox{L}} \tilde{q}_{\tbox{L}})
-
\gY \YR
(\chi_1 \overline{q_{\tbox{R}}} \tilde{q}_{\tbox{R}}^{\ast}
+\chi_1^{\dag} \overline{ q_{\tbox{R}} }^{\dag} \tilde{q}_{\tbox{R}})\nonumber\\
&\ \ -
\gY \YL
(i\chi_2 q_{\tbox{L}} \tilde{q}_{\tbox{L}}^{\ast}
-i\chi_2^{\dag} q^{\dag}_{\tbox{L}} \tilde{q}_{\tbox{L}})
-
\gY \YR
(i\chi_2 \overline{q_{\tbox{R}}} \tilde{q}_{\tbox{R}}^{\ast}
-i\chi_2^{\dag} \overline{ q_{\tbox{R}} }^{\dag} \tilde{q}_{\tbox{R}}),
\end{align}
where we have made a rotation $\chi_2\rightarrow i\chi_2$ so that
its mass appears in the Lagrangian with a positive sign.
Up to a factor of $\sqrt{2}^{-1}$ in the couplings, both $\chi_1^0$ and
$\chi_2^0$ have the interactions similar to the bino of the MSSM.

Integrating out the s-fermions, we have the effective Lagrangian
\begin{align}
\mathcal{L}_{\tbox{eff}}&
=
\frac{\gY^2 Y_L^2}{ M^2_{\tilde{Q}_L} }
\left[
(\overline{\Neu}_1 \PL Q)(\overline{Q} \PR \Neu_1)
+
(\overline{\Neu}_2 \PL Q)(\overline{Q} \PR \Neu_2)
\right.\nonumber\\
&\quad\quad\quad\left.
-
i(\overline{\Neu}_1 \PL Q)(\overline{Q} \PR \Neu_2)
+
i(\overline{\Neu}_2 \PL Q)(\overline{Q} \PR \Neu_1)
\right]\nonumber\\
&+
\frac{\gY^2 Y_R^2}{ M^2_{\tilde{Q}_R} }
\left[
(\overline{\Neu}_1 \PR Q)(\overline{Q} \PL \Neu_1)
+
(\overline{\Neu}_2 \PR Q)(\overline{Q} \PL \Neu_2)\right.\nonumber\\
&\quad\quad\quad\left.
-
i(\overline{\Neu}_1 \PR Q)(\overline{Q} \PL \Neu_2)
+
i(\overline{\Neu}_2 \PR Q)(\overline{Q} \PL \Neu_1)
\right],
\label{eq:Major-eff-Lang}
\end{align}
where $\Neu_i$ (for $i=1,2$) are the four-component Majorana spinor
\begin{align}
\Neu_i\equiv\begin{pmatrix} \chi_i \\ \chi_i^{\dag} \end{pmatrix}.
\end{align}

Although $\chi_1^0$ is the lightest state, it is nearly degenerate with
$\chi_2^0$ with a fractional mass difference of
\begin{align}
\Delta\equiv\frac{M_{\chi_2^0}-M_{\chi_1^0}}{M_{\chi_2^0}}=\frac{\gY^2 v_u v_d}{2M_1\mu},
\end{align}
so the difference in masses between $\chi_2^0$ and $\chi_1^0$
is naturally only a few GeV's.  (For example,
$\Delta=0.04$ for $M_1=75$ GeV, $\mu=500$ GeV, and $\tan\beta=5$.)
The relic density now depends on processes involving $\chi_2^0$,
as $\chi_2^0$ may be abundant when $\chi_1^0$ freezes out \cite{Griest:1990kh}.
In particular,
we have to consider the self- and co-annihilation processes involving $\chi_2^0$ in addition
to the self annihilation of $\chi_1^0$.

Since $\chi^0_{1,2}$ have interactions similar to the Majorana bino
of the MSSM up to a factor in the couplings,
their self annihilation cross sections are of the same form
as Eq.~(\ref{eq:M-Bino-anni})
\begin{align}
\sigma(\chi^0_1\chi^0_1\rightarrow \overline{f}f)
=\frac{\gY^4}{8\pi}\sum\limits_f
N_f
Y_f^4\frac{r_{1f}(1+r_{1f}^2)}{M^2_{\tilde{f}} (1+r_{1f})^4}v,\quad
r_{1f}\equiv \frac{M^2_{\chi^0_1}}{M^2_{\tilde{f}}},
\end{align}
with a similar formula for the annihilation of $\chi_2^0$.
As with the case of the MSSM bino, the self annihilation cross sections
of both $\chi_1^0$ and $\chi_1^0$ are $p$-wave suppressed when thermal-averaged.
The co-annihilation cross section is given by
\begin{align}
\sigma(\chi^0_1\chi^0_2\rightarrow \overline{f}f)v
=\frac{\gY^4}{32\pi}(M_{\chi_1^0}+M_{\chi_2^0})^2
\sum\limits_f
N_f\frac{Y_f}{M^4_{\tilde{f}}}.
\end{align}
Note that this cross section reduces to the annihilation cross section
of the pure Dirac bino in Eq.~(\ref{eq:D-sig}) when $M_{\chi_2^0}=M_{\chi_1^0}$.

We are now ready to calculate the dark matter relic density
taken into account effects of co-annihilation.
We first define
\begin{align}
\sigma_{\tbox{eff}}=\frac{4}{g^2_{\tbox{eff}}}
\left(
\sigma_{\chi_1^0\chi_1^0}
+2\sigma_{\chi_2^0\chi_1^0}(1+\Delta)^{3/2}e^{-x\Delta}
+\sigma_{\chi_2^0\chi_2^0}(1+\Delta)^{3}e^{-2x\Delta}\right),
\end{align}
where $\Delta=(m_{\chi_2^0}-m_{\chi_1^0})/m_{\chi_1^0}$, and
\begin{align}
g_{\tbox{eff}}=2+2(1+\Delta)^{3/2}e^{-x\Delta}.
\end{align}
The relic density of the lightest Majorana
state is now
\begin{align}
\Omega h^2=\frac{x_F}{\sqrt{g_{\ast}}}
\frac{8.7\times 10^{-11}\ \mbox{GeV}^{-2}}{I_a+3I_b/x_F}
\end{align}
where
\begin{align}
I_a= x_F\int\limits_{x_F}^{\infty}\frac{dx}{x^{2}}a_{\tbox{eff}}(x),\quad\mbox{and}\quad
I_b=2x_F\int\limits_{x_F}^{\infty}\frac{dx}{x^{3}}b_{\tbox{eff}}(x).
\end{align}
The functions $a_{\tbox{eff}}$ and $b_{\tbox{eff}}$ are the coefficients of $\sigma_{\tbox{eff}}v$
expanded in $v^2$,
\begin{align}
\sigma_{\tbox{eff}}v=a_{\tbox{eff}}+b_{\tbox{eff}}v^2,
\end{align}
and the freeze-out temperature is solved by the formula similar to
Eq.~(\ref{eq:freeze-out-temp})
\begin{align}
x_F=
\ln\left(
\frac{5}{4}\sqrt{\frac{45}{8}}\frac{g_{\tbox{eff}}}{2\pi^3}
\frac{M_{\chi_1^0} M_{\tbox{Pl}}}{\sqrt{g_{\ast}x_F}}
(a_{\tbox{eff}}+6x_F^{-1}b_{\tbox{eff}})
\right),
\end{align}
with $g_{\tbox{eff}}$, $a_{\tbox{eff}}$, and $b_{\tbox{eff}}$ now
evaluated at the freeze-out temperature.

It is important to note that
the relic density of
pseudo-Dirac bino reduces in the pure Dirac bino limit correctly.
Since the self-annihilation cross sections of $\chi_1^0$ and $\chi_2^0$ are $p$-wave suppressed,
we can make the approximation that
\begin{align}
\sigma_{\tbox{eff}}\simeq\frac{8}{g^2_{\tbox{eff}}}\sigma_{\chi_2^0\chi_1^0}(1+\Delta)^{3/2}e^{-x\Delta},
\end{align}
which is valid as long as the exponential Boltzmann suppression $e^{-x_F\Delta}$ is much larger than
$p$-wave suppression of the self annihilation $x^{-1}_F$.  (For example, for $\mu=500$ GeV, $\tan\beta=5$, and
$x_F=20$, we have $e^{-x_F\Delta}\sim 0.66$ while $x_F^{-1}=0.05$, so the approximation
is valid.)
In the Dirac bino limit of $\Delta\rightarrow 0$, the effective cross section
$\sigma_{\tbox{eff}}v$ is half of the annihilation of pure Dirac bino in Eq.~(\ref{eq:D-sig}) because of
the factor $g_1 g_2 g^{-2}_{\tbox{eff}}$ in $\sigma_{\tbox{eff}}v$ approaches $\tfrac{1}{2}$, naively leading to
a relic density that is twice as large as the pure Dirac bino.
However, in the case of the pure Dirac bino,
there is an additional factor of 2 in its relic density to account for both the particle and
antiparticle, and the relic density of pseudo-Dirac bino approaches that
of the pure-Dirac bino correctly.

We can also find the leading dependence of the relic density of the pseudo-Dirac bino
on the splitting in mass $\Delta$.
For small $\Delta$ (such that $e^{x_F\Delta}\gg x_F^{-1}$), where the main annihilation mode of
the $\chi_2^0-\chi_1^0$ system
is the co-annihilation mode, the annihilating cross section
is suppressed compared to the annihilation in the limit of
pure Dirac bino by
a factor of
\begin{align}
\frac{I_a}{\langle\sigma(\overline{D}D\rightarrow \overline{f}f)v\rangle}
=2\frac{e^{-x_F\Delta}}{\left[1+(1+\Delta)^{3/2}e^{-x_F\Delta}\right]^2}
\left(1+\tfrac{5}{2}\Delta+\mathcal{O}\left(\Delta^2\right)\right),
\end{align}
when we expand $I_a$ in $\Delta$.
The relic density of the pseudo-Dirac bino increases correspondingly by
(taking into account the factor of 2 in the relic density of pure
Dirac bino)
\begin{align}
\frac{\Omega_{\mathcal{N}}}{\Omega_{D}}=
\frac{1}{4}\left(e^{x_F\Delta}\right)\left[1+(1+\Delta)^{3/2}e^{-x\Delta}\right]^2
\left(1-\tfrac{5}{2}\Delta+\mathcal{O}\left(\Delta^2\right)\right),
\end{align}
and we can explicitly see the relic density of pseudo-Dirac bino reduces correctly in the pure Dirac bino limit
($\Delta\rightarrow 0$).

\begin{figure}[htb]
\begin{center}
\includegraphics[width=4in]{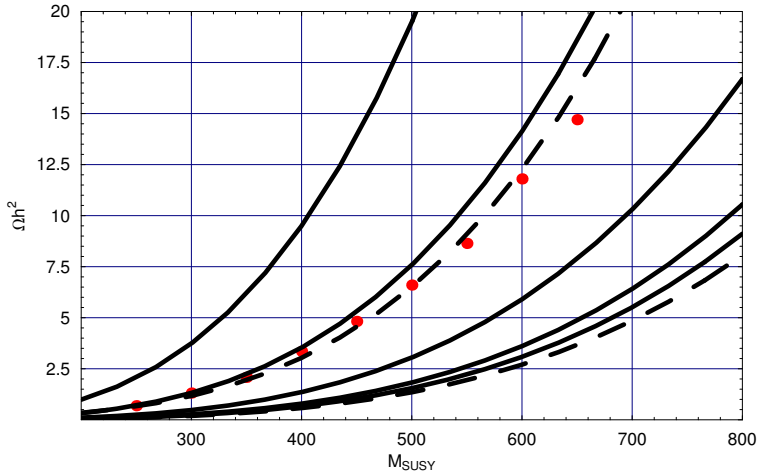}
\caption{
The dashed lines are those in Figure \ref{fig:DBinoPlot}.
The solid lines, from bottom to top, correspond to the relic densities of
the pseudo-Dirac plot for $\Delta=0.01, 0.02, 0.05, 0.10, $ and $0.15$.}
\label{fig:PBinoPlot}
\end{center}
\end{figure}

In Figure \ref{fig:PBinoPlot}, we plot the relic density of
pseudo-Dirac bino as a function of a common scalar soft mass
$M_{\tbox{SUSY}}$ for several values of $\Delta$.  We
see that, even for $\Delta=0.05$, the relic density of the
pseudo-Dirac bino is still less than the Majorana bino
by about a factor of 2.  For $\Delta=0.10$, the relic density of
the pseudo-Dirac bino is about the same, though slightly larger, as that of the MSSM bino.
For $\Delta=0.15$, the pseudo-Dirac bino relic density is larger than
that of the MSSM bino by about a factor of 3, signalling the decreasing effects of co-annihilation
and the weaker interactions between matter and the lighter pseudo-Dirac bino state $\chi_1^0$
compared to the MSSM bino.

\subsection{Direct detection in the pure Dirac bino limit}
The direct detection experiments \cite{cdms}\cite{Angle:2007uj} measure
recoils of heavy nuclei from
interactions with dark matter.  The recoil energies are of the scale
of tens of keV, and the bounds are expressed in terms of
elastic cross sections between dark matter and the nucleon.
The most stringent bounds set by these experiments
come from the spin-independent interactions
between dark matter and the nuclei, and
it is only those interactions that we consider
for the pure Dirac bino.

To compute the elastic cross section between
dark matter and the nucleon,
we re-write the effective interaction of Eq.~(\ref{eq:D-eff2})
as vector and axial-vector interactions
\begin{align}
\mathcal{L}_{\tbox{eff}}
&=
\left[
a_{\tbox{L}}(\overline{D}^{\prime} \gamma^{\mu} \PL D^{\prime})
+
a_{\tbox{R}}(D\gamma^{\mu}\PR D)
\right]
\overline{Q}\gamma_{\mu}Q
\nonumber\\
&\quad+
\left[
a_{\tbox{L}}(\overline{D}^{\prime} \gamma^{\mu} \PL D^{\prime})
-
a_{\tbox{R}}(D\gamma^{\mu}\PR D)
\right]
\overline{Q}\gamma_{\mu}\gamma^5 Q,
\label{eq:D-eff3}
\end{align}
where
\begin{align}
a_{\tbox{L,R}}=\frac{\gY^2 Y^2_{ \tbox{L,R} } }{2M^2_{ \tilde{Q}_{ \tbox{L,R} } }}.
\end{align}
As vector contributions of the quarks in the nucleus add coherently, we
can express the cross section between Dirac bino and a nucleus $N(Z,A)$ as
\cite{Jungman:1995df}
\begin{align}
\sigma^N_{\tbox{vec}}
=
\frac{b_N^2}{\pi}\frac{M_1^2 m_N^2 }{(M_1+m_N)^2},
\label{eq:vec-nucleus}
\end{align}
with $b_N=Z b_p+(A-Z)b_n$, $b_p=2b_u+b_d$, $b_n=b_u+2b_d$, and
\begin{align}
b_u&=\frac{1}{2}( a_{u_{\tbox{L}}}+a_{u_{\tbox{R}}} )
=\frac{\gY^2}{4}
\left(
\frac{1}{36M^2_{\tilde{u}_{\tbox{L}}}}
+
\frac{4}{9M^2_{\tilde{u}_{\tbox{R}}}}
\right),
\\
b_d&= \frac{1}{2}( a_{d_{\tbox{L}}}+a_{d_{\tbox{R}}} )
=\frac{\gY^2}{4}
\left(
\frac{1}{36M^2_{\tilde{d}_{\tbox{L}}}}
+
\frac{1}{9M^2_{\tilde{d}_{\tbox{R}}}}
\right).
\end{align}

The experimental bounds are expressed in the bino-nucleon cross section $\sigma_n$ that
is related to the bino-nucleus cross section $\sigma_N$ by
\begin{align}
\sigma^{n}_{\tbox{vec}}
=\frac{M_n}{M_N}\frac{(M_1+M_N)^2}{(M_1+M_n)^2}\frac{\sigma^N_{\tbox{vec}}}{A^2}.
\end{align}
In the simplified case where all the sfermion masses are degenerate with a common
mass $M_{\tbox{SUSY}}$, for the $^{73}$Ge detector used in CDMS II \cite{cdms}, the bino-nucleon cross section
is
\begin{align}
\sigma^{n}_{\tbox{vec}}=(8.6\times 10^{-39})\left(\frac{500\ \mbox{GeV}}{M_{\tbox{SUSY}}}\right)^4
\ \mbox{cm}^{2}.
\end{align}
This is well above the upper-bound of $2\times 10^{-43}$ cm$^2$ set by CDMS II for dark
matter with mass on the order of 100 GeV, and the limit of pure Dirac bino with mass of the scale of 100 GeV
is ruled out as dark matter.

\subsection{Direct detection of pseudo-Dirac bino}
As stated in the Introduction, as long as the splitting between the two states of
the pseudo-Dirac bino is larger than 10s of keVs, the direct detection experiments
are only sensitive to the lighter state $\chi_1^0$.
Since $\chi_1^0$ in our approximation behaves exactly as the MSSM bino up to
a scaled coupling, the direct detection bounds are similar to the case of the MSSM bino.
The direct detection rates of the MSSM neutralino has been studied extensively in the literature
\cite{Griest:1988ma}\cite{Jungman:1995df}\cite{Bottino:1996eu}\cite{Choi:2000kh}\cite{Nihei:2004bc}.
In particular, being a Majorana particle, there is no longer a vector interaction
with the quarks, and the resulting $\chi_1^0$-nucleon cross section is much smaller.
Here we will simply state the results from the literature for the direct detection
rates for the MSSM bino $\tilde{B}$
(for the pseudo-Dirac bino $\chi_1^0$,
simply make the replacement $\gY\rightarrow\sqrt{2}^{-1}\gY$).  Our presentation
here is mainly based on Reference \cite{Choi:2000kh}.

The four-fermion effective Lagrangian for the bino $\tilde{B}$ is given by
\begin{align}
\mathcal{L}=\frac{\gY^2}{2}(\overline{\tilde{B}}\gamma^{\mu}\gamma_5\tilde{B})
\left[\frac{\YL^2}{M^2_{\tilde{Q}_L}}(\overline{Q}\gamma_{\mu}\PR Q)
-\frac{\YR^2}{M^2_{\tilde{Q}_R}}(\overline{Q}\gamma_{\mu}\PL Q)\right],
\end{align}
since $\overline{\tilde{B}}\gamma_{\mu}\tilde{B}=0$, there are only axial-vector
interactions with the coefficients
\begin{align}
A_Q=\frac{\gY^2}{4}\left(\frac{\YL^2}{M^2_{\tilde{Q}_L}}
+\frac{\YR^2}{M^2_{\tilde{Q}_R}}\right).
\end{align}

The evaluation of the elastic cross section will now
require the matrix elements
\begin{align}
\langle n|\overline{Q}\gamma_{\mu}\gamma_5Q|n\rangle=2s_{\mu}^{n}
\Delta_Q^{n},
\end{align}
where $s_{\mu}^{n}$ is the spin of the nucleon $n$,
and $\Delta_Q^{n}$ (extracted from experiments) is the fraction of nucleon spin carried by
quark $Q$.  The experimental values are \cite{Adams:1994zd}
\begin{align}
\Delta_u^p=0.77,&\quad\Delta_d^p=-0.38,\quad\Delta_s^p=-0.09,\quad\nonumber\\
\Delta_u^n=-0.38,&\quad\Delta_d^n=0.77,\quad\Delta_s^n=-0.09.
\end{align}

The elastic cross section is then
\begin{align}
\sigma^N_{\tbox{axial-vec}}=
\frac{16}{\pi}
\frac{M^2_{\chi_1^0}M_N^2}{(M^2_{\chi_1^0}+M_N)^2}
\frac{J+1}{J}
\left(
\langle S_p\rangle\sum\limits_{u,d,s}(A_Q\Delta_Q^p)
+
\langle S_n\rangle\sum\limits_{u,d,s}(A_Q\Delta_Q^n)
\right)^2,
\end{align}
where $J$ is the spin of the nucleus, $\langle S_{p,n}\rangle=\langle N|S_{p,n}|N\rangle$ are the
expectation values of the spin content of the proton and neutron
groups in the nucleus, respectively.
Their values values $\langle S_{p,n}\rangle$ for
$^{73}$Ge are given by the shell model
as \cite{Ressell:1993qm}
\begin{align}
\langle S_{p}\rangle_{\tbox{Ge}}=+0.011,\quad
\langle S_{n}\rangle_{\tbox{Ge}}=-0.491.
\end{align}

For $^{73}$Ge ($J=\tfrac{9}{2}$), $M_{\chi_1^0}=75$ GeV,
and a common squark mass of $M_{\tbox{SUSY}}$, the
spin-dependent cross section is then
\begin{align}
\sigma^N_{\tbox{axial-vec}}=1.0\times 10^{-42}\left(\frac{500\ \mbox{GeV}}{M_{\tbox{SUSY}}}\right)^4
\ \mbox{cm}^{2},
\end{align}
which is consistent with the CDMS II upper bounds of $1\times 10^{-38}$ cm$^2$.
It should be noted, however, that Higgsino components of $\chi_1^0$ that we ignore here
may change the direct detection rates significantly.  The Higgsinos have scalar
interactions with nucleus, which are coherent and proportional to the nucleus mass.
If Higgsino composition of $\chi_1^0$ are significant,the spin-independent cross section
may overwhelm the spin-dependent cross section.  We will leave this for future work.

\section{CONCLUSIONS}
\label{sec:Conc}
In this paper we have calculated the relic density and direct detection rates
for pseudo-Dirac bino, which arises naturally as dark matter in supersymmetric
models with $D$-type SUSY-breaking.  Although we have performed these calculations
in some very simple limits, our results are nonetheless interesting.
For small mass splitting between the two pseudo-Dirac bino states
(of a few percent
in the fractional difference in masses), the
relic density of pseudo-Dirac bino is closer to WMAP observations
compared to the MSSM bino, while its direct detection rate is smaller
than the MSSM bino by a factor of 4.  The reduced relic density
of the pseudo-Dirac bino implies that there may be less fine-tuning
of the D-type SUSY-breaking
spectra to achieve a dark matter relic density consistent with observations.

As with the rich phenomenology of the neutralino sector of the MSSM,
relaxing any of the assumptions of this study can lead to significantly
different conclusions.  In particular, it would be interesting
to include annihilation to the gauge and Higgs bosons.  Also, the relative
signs between the various mass parameters can be important, as well as
the wino/Higgsino mixture of $\chi_1^0$.  In addition, although
qualitatively there may be less fine-tuning to achieve observed
relic density, it is important to quantify the degree of fine-tuning
and compare it with the MSSM.  We leave these open projects for future work.

\section{ACKNOWLEDGEMENTS}
I would like to thank Professor Markus Luty for initiating this project and the many useful
discussions.
I also thank Professors Zackaria Chacko and Rabindra Mohapatra for helpful comments
and discussions.  I thanks to Nick Setzer, Sogee Spinner, and Haibo Yu for very useful
comments on specific aspects of many calculations.
I would also like to thank the High Energy Group of
Michigan State University
for its hospitality during part of this work.
I use CalcHEP \cite{Pukhov:2004ca} to check parts of the calculations of
this work, and thank Neil Christensen for his help with CalcHEP.
This work is supported by NSF Grant PHY-0354401.


\begin{thebibliography}{99}

\bibitem{Griest:1988ma}
  K.~Griest,
  Phys.\ Rev.\  D {\bf 38}, 2357 (1988)
  [Erratum-ibid.\  D {\bf 39}, 3802 (1989)].

\bibitem{Griest:1989zh}
  K.~Griest, M.~Kamionkowski and M.~S.~Turner,
  Phys.\ Rev.\  D {\bf 41}, 3565 (1990).

\bibitem{Jungman:1995df}
  G.~Jungman, M.~Kamionkowski and K.~Griest,
  Phys.\ Rept.\  {\bf 267}, 195 (1996)
  [hep-ph/9506380]

\bibitem{Bottino:1996eu}
  A.~Bottino, F.~Donato, G.~Mignola, S.~Scopel, P.~Belli and A.~Incicchitti,
  Phys.\ Lett.\  B {\bf 402}, 113 (1997)
  [hep-ph/9612451].

\bibitem{Choi:2000kh}
  S.~Y.~Choi, S.~C.~Park, J.~H.~Jang and H.~S.~Song,
  Phys.\ Rev.\  D {\bf 64}, 015006 (2001)
  [hep-ph/0012370].

\bibitem{Nihei:2002ij}
  T.~Nihei, L.~Roszkowski and R.~Ruiz de Austri,
  JHEP {\bf 0203}, 031 (2002)
  [hep-ph/0202009].


\bibitem{Nihei:2004bc}
  T.~Nihei and M.~Sasagawa,
  Phys.\ Rev.\  D {\bf 70}, 055011 (2004)
  [Erratum-ibid.\  D {\bf 70}, 079901 (2004)]
  [hep-ph/0404100].

\bibitem{wmap} D.~N.~Spergel et al.,
[astro-ph/0603449].

\bibitem{EllisKingRoberts}
  J.~R.~Ellis and K.~A.~Olive,
  Phys.\ Lett.\  B {\bf 514}, 114 (2001)
  [hep-ph/0105004].
\\
  S.~F.~King and J.~P.~Roberts,
  JHEP {\bf 0609}, 036 (2006)
  [hep-ph/0603095].
\\
  S.~F.~King and J.~P.~Roberts,
  JHEP {\bf 0701}, 024 (2007)
  [hep-ph/0608135].
\\
  S.~F.~King, J.~P.~Roberts and D.~P.~Roy,
  arXiv:0705.4219 [hep-ph].


\bibitem{Arkani-Hamed:2006mb}
  N.~Arkani-Hamed, A.~Delgado and G.~F.~Giudice,
  Nucl.\ Phys.\  B {\bf 741}, 108 (2006)
  [hep-ph/0601041].

\bibitem{acd}
  T.~Appelquist, H.~C.~Cheng and B.~A.~Dobrescu,
  Phys.\ Rev.\  D {\bf 64}, 035002 (2001)
  [hep-ph/0012100].

\bibitem{Servant:2002hb}
  G.~Servant and T.~M.~P.~Tait,
  New J.\ Phys.\  {\bf 4}, 99 (2002)
  [hep-ph/0209262].



\bibitem{cdms}
  D.~S.~Akerib {\it et al.}  [CDMS Collaboration],
Phys.\ Rev.\ Lett.{\bf 96} (2006) 011302
[astro-ph/0509259].
\\
  D.~S.~Akerib {\it et al.}  [CDMS Collaboration],
  Phys.\ Rev.\  D {\bf 73}, 011102 (2006)
  [astro-ph/0509269].

\bibitem{Angle:2007uj}
  J.~Angle {\it et al.}  [XENON Collaboration],
  arXiv:0706.0039 [astro-ph].

\bibitem{KH}
  K.~Hsieh, R.~N.~Mohapatra and S.~Nasri,
  JHEP {\bf 0612}, 067 (2006)
  [hep-ph/0610155].
\\
  K.~Hsieh, R.~N.~Mohapatra and S.~Nasri,
  Phys.\ Rev.\  D {\bf 74}, 066004 (2006)
  [hep-ph/0604154].



\bibitem{Belanger:2007dx}
  G.~Belanger, A.~Pukhov and G.~Servant,
  arXiv:0706.0526 [hep-ph].


%
\bibitem{DGMSB2}
  P.~J.~Fox, A.~E.~Nelson and N.~Weiner,
  JHEP {\bf 0208}, 035 (2002)
  [hep-ph/0206096].
\\
  I.~Antoniadis, K.~Benakli, A.~Delgado and M.~Quiros,
  [hep-ph/0610265].
\\
  L.~M.~Carpenter, P.~J.~Fox and D.~E.~Kaplan,
  [hep-ph/0503093].

\bibitem{Hall:1997ah}
  L.~J.~Hall, T.~Moroi and H.~Murayama,
  Phys.\ Lett.\  B {\bf 424}, 305 (1998)
  [hep-ph/9712515].


\bibitem{lep2}
  A.~Heister {\it et al.}  [ALEPH Collaboration],
  Phys.\ Lett.\  B {\bf 526}, 191 (2002)
  [hep-ex/0201014].
\\
  J.~Abdallah {\it et al.}  [DELPHI Collaboration],
  Eur.\ Phys.\ J.\  C {\bf 32}, 145 (2004)
  [hep-ex/0303013].
\\
  M.~Acciarri {\it et al.}  [L3 Collaboration],
  Phys.\ Lett.\  B {\bf 519}, 33 (2001)
  [hep-ex/0102025].
\\
  G.~Abbiendi {\it et al.}  [OPAL Collaboration],
  Eur.\ Phys.\ J.\  C {\bf 26}, 479 (2003)
  [hep-ex/0209078].
\\
  R.~Barate {\it et al.}  [LEP Working Group for Higgs boson searches],
  Phys.\ Lett.\  B {\bf 565}, 61 (2003)
  [hep-ex/0306033].


\bibitem{KolbTurner}
  E.~W.~Kolb and M.~S.~Turner,
  Front.\ Phys.\  {\bf 69}, 1 (1990).

\bibitem{Srednicki:1988ce}
  M.~Srednicki, R.~Watkins and K.~A.~Olive,
  Nucl.\ Phys.\  B {\bf 310}, 693 (1988).


\bibitem{micromegas}
  G.~Belanger, F.~Boudjema, A.~Pukhov and A.~Semenov,
  Comput.\ Phys.\ Commun.\  {\bf 176}, 367 (2007)
  [hep-ph/0607059].

\bibitem{Griest:1990kh}
  K.~Griest and D.~Seckel,
  Phys.\ Rev.\  D {\bf 43}, 3191 (1991).

\bibitem{Adams:1994zd}
  D.~Adams {\it et al.}  [Spin Muon Collaboration (SMC)],
  Phys.\ Lett.\  B {\bf 329}, 399 (1994)
  [hep-ph/9404270].

\bibitem{Ressell:1993qm}
  M.~T.~Ressell, M.~B.~Aufderheide, S.~D.~Bloom, K.~Griest, G.~J.~Mathews and D.~A.~Resler,
  Phys.\ Rev.\  D {\bf 48}, 5519 (1993).

\bibitem{Pukhov:2004ca}
  A.~Pukhov,
  arXiv:hep-ph/0412191.


\end{thebibliography}
\end{document}